\documentstyle[12pt]{article}
\def\be{\begin{equation}}
\def\ee{\end{equation}}
\newcommand{\bea}{\begin{eqnarray}}
\newcommand{\eea}{\end{eqnarray}}
\def\bar{\widetilde}
\def\H{{\cal H}}
\def\singlespacing{\baselineskip=12pt}
\def\doublespacing{\baselineskip=24pt}
\def\f#1{{\mbox{\boldmath $#1$}}}
\def\factor{\lambda}
\def\F{{F}}
\def\Fh{{F}}
\def\Fb{{P}}
\def\Phib{U}

\def\d{{{\mbox{\boldmath $d$}}}}

\def\dx{\d{x}}
\def\dt{\d{t}}

\def\bdry{{\partial{\cal M}}}
\def\ba{{I_{\partial{\cal M}}}}        
\def\C{{\cal C}}
\begin{document}
\noindent
\pagestyle{empty}
December 9, 1994 \\
\def\K{{\bf K}}
\def\inv{\f\omega}
\def\O{{\cal O}}
\def\Ls{\pounds}
\bigskip
\bigskip
\doublespacing
\font\mainh=cmbx10 scaled \magstep1
\baselineskip=24.9truept plus 2pt
\def\J{{\bf J}}
\def\frac#1#2{{\textstyle {#1\over #2}}}
\medskip
\begin{center}
\begin{large}
{\mainh The Cosmological Probability Density Function for
Bianchi Class A Models in Quantum Supergravity}\\
\end{large}
\bigskip
\bigskip
HUGH LUCKOCK and CHRIS OLIWA \\
\smallskip
School of Mathematics and Statistics\\
University of Sydney\\
NSW 2006\\
Australia\\
\end{center}
\bigskip
\bigskip
\bigskip
\leftline{\bf Abstract}\smallskip\noindent
Nicolai's theorem suggests a simple stochastic interpetation
for supersymmetric Euclidean quantum theories, without
requiring any inner product to be defined on the space
of states. In order to apply this idea to supergravity, we first
reduce to a one-dimensional theory with local supersymmetry by the
imposition of homogeneity conditions. We then make the supersymmetry
rigid by imposing gauge conditions, and quantise to obtain the
evolution equation for a time-dependent wave function. Owing
to the inclusion of a certain boundary term in the classical
action, and a careful treatment of the initial conditions,
the evolution equation has the form of a Fokker-Planck equation.
Of particular interest is the static solution, as this satisfies
all the standard quantum constraints. This is naturally interpreted
as a cosmological probability density function, and is found to
coincide with the square of the magnitude of the conventional wave
function for the wormhole state.

\bigskip
\leftline{PACS numbers: 98.80.H , 04.65}

\newpage
\pagestyle{plain}

\pagenumbering{arabic}

\noindent

\leftline{\bf Introduction}

Supersymmetric theories enjoy a number of appealing properties,
many of which are particularly attractive in the context of quantum
cosmology. For example, supersymmetry leads to the replacement
of second-order constraints by first-order constraints, which
are both easier to solve and more restrictive.

A feature of supersymmetry which has not yet been fully exploited
in quantum cosmology arises from a theorem proved by Nicolai in
1980 \cite{Nic}. In essence, this theorem states that any Euclidean
theory with rigid supersymmetry can be converted into a free bosonic
theory by integrating out the fermions and transforming the
bosonic variables. The new variables then have the stochastic properties
of white noise, and the transformation which relates them to the
original bosonic variables (the ``Nicolai map") is typically a
stochastic differential equation of the Langevin type. This leads
to a natural probabilistic intepretation of the quantum theory,
without requiring the introduction of any inner product on the
space of states

Recent work has shown how these ideas can be applied to quantum
cosmological models with $N=2$ supersymmetry \cite{GL1}. However
these models are somewhat artificial, as they are constructed
by modifying purely bosonic minisuperspace models. Consequently,
this kind of supersymmetry does not necessarily reflect any
underlying physical symmetry of the full theory.

In this paper we overcome this limitation by applying similar
ideas to supergravity. Although the full theory is not amenable
to our approach (due to the breaking of supersymmetry by the
imposition of initial or final conditions on the fields),
we are able to make progress by focussing on homogeneous field
configurations in Bianchi class A spacetimes. This reduction
leads to a one-dimensional model with reparametrisation
invariance and local $N=4$ supersymmetry.

Before invoking Nicolai's theorem, it is necessary to convert
the local supersymmetry to a rigid supersymmetry, and to fix
the reparametrisation invariance. Both these ends are easily
achieved by imposing various gauge conditions on the classical
theory. This procedure also results in one of the most novel
and useful features of our approach; a quantum theory which
evolves with respect to clock time.

Though unorthodox, it should be noted that there is nothing
unreasonable about a cosmological wave function which evolves
with respect to the proper time measured by a family of observers
distributed through the Universe. Of course, this evolution depends
completely upon the selection of these observers; or equivalently,
upon the choice of gauge.

Since the theory now has rigid supersymmetry, it must admit a Nicolai
map. This will be a type of Langevin equation governing the dynamics of
the bosonic variables. In principle, its explicit form could be
obtained by integrating out the fermionic variables in the
path integral. However, we are primarily interested in the
probabilistic interpretation of the theory rather than the
Nicolai maps \it per se \rm. For this reason, we bypass
the path integral formulation and use instead canonical
quantisation. This leads directly to a type of Fokker-Planck
equation governing the Euclidean evolution of the wave
function.

We wish to emphasise that a plausible probabilistic interpretation
of the quantum theory is obtained without introducing an inner
product or appealing to the concept of a Hilbert space. In
this sense our approach is fundamentally different from
conventional quantum mechanics, even though we ultimately
obtain the same expression for the static probability density
function as can be obtained by a more conventional approach.
\bigbreak

\leftline{\bf Classical Supergravity}

The basic fields in supergravity are the tetrad $e^{AA'}{}_\mu$ and the
spin $3/2$ Rarita-Schwinger fields $\psi^A{}_\mu$ and $\bar\psi^{A'}{}_\mu$.
(We use the two component notation described in \cite{D1}, in which
spinor indices $A,A'$ take the values 0 and 1, and are raised or lowered
with the antisymmetric quantities $\epsilon^{AB}$, $\epsilon_{AB}$.)
The action on an unbounded region of spacetime ${\cal M}$ has the form
\begin{equation}
I =
\int_{\cal M} d^4\!x\left[{1\over
2\kappa^2}eR+\frac{1}{2}\epsilon^{\mu\nu\rho\sigma}
\left(\bar\psi^{A'}{}_\mu e_{AA'\nu} D_{\rho}\psi^A{}_\sigma
-\psi^A{}_\mu e_{AA'\nu} D_{\rho}\bar\psi^A{}_\sigma\right)\right]
                                  \label{eq:supgr_action}
\ee
where $e$ is the determinant of the tetrad, $R$ is the scalar curvature,
and $D_\mu$ denotes the covariant spinor derivative. If ${\cal M}$
has boundary $\bdry$, then the action $I$ must also include the boundary term
$\kappa^{-2} \int_{\bdry} K h^{1\over 2} d^3x$, where $K$ is the
trace of the second fundamental form on $\bdry$ and $h^{1\over 2}d^3x$ is
the induced volume element on $\bdry$ \cite{GH}.
Requiring the total action to be stationary with respect to independent
variation of the tetrad and the connection then leads to Einstein's equations
and to the ``tetrad postulate" (which relates the tetrad and connection)
\cite{vN}.

When the spacetime has a Lorentzian ($-+++$) signature, the reality of the
action
demands that the components $e^{AA'}{}_\mu$ of the tetrad are Hermitian while
the left-handed spinors $\bar\psi^{A'}{}_\mu$ are the Hermitian conjugates of
the right-handed spinors $\psi^A{}_\mu$. However, not wanting to restrict
ourselves to Lorentzian spacetimes, we will think of all these fields as being
unconstrained.

As well as coordinate and Lorentz transformations, the supergravity action
is invariant (up to boundary terms) under the
local supersymmetry transformations
\be
\delta e^{AA'}{}_\mu = -i\kappa(\epsilon^A\bar\psi^{A'}{}_\mu +
                       \bar\epsilon^{A'}\psi^A{}_\mu), \ \ \
\delta\psi^A{}_\mu  =  2\kappa^{-1}D_\mu\epsilon^A,\ \ \
\delta\bar\psi^{A'}{}_\mu=
2\kappa^{-1}D_\mu{\bar{\epsilon}}^{A'}.\label{eq:transf}
\ee
where the spinors $\epsilon^A, \bar\epsilon^{A'}$ have anticommuting
Grassmann components.

It was shown in \cite{L1} that, when initial or final conditions are
imposed, the validity of Nicolai's theorem depends on the existence of
an unbroken subalgebra of supersymmetry generators. The first step
in our program is therefore to identify a (non-trivial) subalgebra
of supersymmetry generators which preserves the hypersurfaces
on which we will specify the initial and final data. In fact, the subalgebra
of left-handed supersymmetry generators obtained by setting
$\epsilon=0$ in (\ref{eq:transf}) is adequate for this purpose, since
the anticommutator of two such generators is always zero.
(The subalgebra of right-handed supersymmetry generators
obtained by setting $\bar\epsilon=0$ would do equally well.)

We cannot proceed immediately with our program, as the supersymmetric
variation of the action gives rise to boundary terms which invalidate
Nicolai's theorem \cite{L1}. We therefore seek a boundary correction
which restores the invariance of the supergravity action under the
chosen subalgebra.

In fact D'Eath has shown that there is no such boundary correction
for the full supergravity theory, which is therefore not amenable to
our approach \cite{D2}. However, it turns out that we \it can \rm complete
the program outlined above if we restrict our attention to spatially
homogeneous Bianchi class A models.

In Bianchi class A models, spacetime is parametrised by a global timelike
coordinate $t=x^0$ and three locally defined spacelike coordinates $x^i \
(i=1,2,3)$; there are also three 1-forms
$\inv^p=\omega^p{}_i \dx^i$ ($p=1,2,3$) satisfying
\be
\d\inv^p = {1\over 2} m^{ps} \epsilon_{sqr} \inv^q\wedge\inv^r
\label{eq:hom_cond}
\ee
where $m^{ps}= m^{(ps)}$ is some constant symmetric matrix \cite{RS}.
The tetrad and the spin $3/2$ field are required to be spatially
constant in the sense that
\be
e^{AA'}{}_i=  e^{AA'}{}_p(t) \omega^p{}_i, \ \ \ \ \ \
\psi^A{}_i= \psi^A{}_p(t) \omega^p{}_i,\ \ \ \ \
\bar\psi^{A'}{}_i= \bar\psi^{A'}{}_p(t) \omega^p{}_i
\ee
with $e^{AA'}{}_p(t)$, $\psi^A{}_p(t)$, $\bar\psi^{A'}{}_p(t)$ and
$e^{AA'}{}_0(t)$, $\psi^A{}_0(t)$, $\bar\psi^{A'}{}_0(t)$ being independent
of the spatial coordinates $x^i$. The 3-metric on each spatial hypersurface
$\Sigma(t)$
of constant $t$ then has the form
\be
h_{pq}(t) \inv^p\otimes\inv^q\ \ \ \ \hbox{where}\ \ \
h_{pq} = - e^{AA'}{}_p e_{AA'q} .
\ee

Having imposed a homogeneous ansatz on the fields, we now need a boundary
correction which restores the invariance of the action under the chosen
subalgebra. Indeed, the action is found to be invariant under the
left-handed subalgebra if one adds the boundary correction \cite{Oliwa}
\be
\ba = \int_{\partial{\cal M}}
\left( i\kappa^{-2} m^{pq}h_{pq}
-{1\over 2} \epsilon^{pqr}\bar\psi^{A'}{}_p
\psi^A{}_q e_{AA'r}  \right) \inv^1\wedge\inv^2\wedge\inv^3
\ee
where the spacetime boundary $\partial{\cal M}$ is assumed to
consist of two disjoint hypersurfaces $\Sigma(t_{initial})$
and $\Sigma(t_{final})$
on which initial and final data are to be specified.
Alternatively, the action can be made invariant
under the right-handed subalgebra by subtracting $\ba $.
We thus obtain two new forms of the classical action
\begin{equation}
I_\pm \equiv I \pm  \ba
\end{equation}
with $I_+$ invariant under the left-handed subalgebra,
and $I_-$ invariant under the right-handed subalgebra.

Note that the boundary corrections are not real, and so their inclusion
corresponds to a non-unitary transformation in the (Lorentzian) quantum
theory. However we are ultimately concerned with the Euclidean
theory, where unitarity is not required \cite{GL1}.

Before quantisation, we consider briefly the canonical form of the
classical theory. We take the dynamical variables as
$e^{AA'}{}_p$, $\psi^A{}_p$, $\bar\psi^{A'}{}_p$ and
$p_{AA'}{}^p$, the latter denoting the momenta conjugate to
$e^{AA'}{}_p$ in the original theory defined by eq. (\ref{eq:supgr_action}).
There are second-class constraints relating the spinor fields to
their own conjugate momenta, and so the latter need not be
treated as independent fields.

The addition of the boundary corrections $\pm\ba$ to the original
action suggests that we should also define new momenta
\bea
p^\pm{}_{AA'}{}^p & \equiv & p_{AA'}{}^p
\pm  {\partial \ba\over \partial e^{AA'}{}_p} \\
 &= &  p_{AA'}{}^p \mp   2i \kappa^{-2} \sigma m^{pq}e_{AA'q}
\mp  {1\over 2} \sigma\epsilon^{pqr}\bar\psi_{A'q}\psi_{Ar} \label{eq:new_mom}
\eea
where $\sigma$ is the integral of the 3-form $\inv^1\wedge\inv^2\wedge\inv^3$
over the spatial hypersurface $\Sigma(t)$.
Then $p^+{}_{AA'}{}^p$ are the momenta conjugate to $e^{AA'}{}_p$ in the
version of the theory invariant under the left-handed subalgebra, while
$p^-{}_{AA'}{}^p$ are the momenta in the version invariant under the
right-handed subalgebra.

Because there are second-class constraints, the usual Poisson brackets
are replaced by Dirac brackets and their fermionic
generalisations \cite{D1,Dir,Cas}. These are very simple when written
in terms of the new momenta $p^\pm{}_{AA'}{}^p$. In particular, one finds
that
\be
[p^+{}_{AA'}{}^p, p^+{}_{BB'}{}^q]^\ast  = 0\ \ \ \ \ \
[p^-{}_{AA'}{}^p, p^-{}_{BB'}{}^q]^\ast  = 0    \label{eq:Dir_1}
\ee
\be
[p^+{}_{AA'}{}^p , \psi^B{}_q]^\ast   = 0 \ \ \ \ \ \
[p^-{}_{AA'}{}^p , \bar\psi^{B'}{}_q]^\ast  = 0. \label{eq:Dir_2}
\ee
The vanishing of these brackets is a pleasant surprise; the
corresponding brackets involving the usual momenta $p_{AA'}{}^p$
do \it not \rm vanish.

We also find that
\bea
[ e^{AA'}{}_p\, ,\psi^B{}_q ]^\ast                   & = &  0 \\
{}[ e^{AA'}{}_p\, ,\bar\psi^{B'}{}_q ]^\ast            & = &  0 \\
{}[ \psi^A{}_p\, ,\psi^B{}_q ]^\ast                    & = &  0 \\
{}[ \bar\psi^{A'}{}_p,\bar\psi^{B'}{}_q ]^\ast         & = &  0 \\
{} [e^{AA'}{}_{p},p^{\pm}{}_{BB'}{}^q ]^\ast           & = &
\epsilon^A{}_B\epsilon^{A'}{}_{B'}\delta^q_p\, \label{eq:Dir_3} \\
{}[\psi^A{}_p,\bar\psi^{A'}{}_q ]^\ast                 & = &
 -{1\over\sigma}D^{AA'}{}_{pq}\   \label{eq:Dir_4}\\
{}[p^+{}_{AA'}{}^p,\bar\psi^{B'}{}_q ]^\ast            & = &
\epsilon^{pqr} D_A{}^{B'}{}_{rs} \bar\psi_{A'q}\\
{}[p^-{}_{AA'}{}^p,\psi^B{}_q ]^\ast            & = &
\epsilon^{pqr} D^B{}_{A'sq}\psi_{Ar}  \label{eq:Dir_last}
\eea
where
\be
D^{AA'}{}_{pq} = -2i (\det [h_{pq}] )^{-1/2} e^{AB'}{}_q e_{BB'p} n^{BA'},
\ee
and $n^{AA'}=e^{AA'}{}_\mu n^\mu$ is the spinor version of the future-pointing
unit vector normal to the spatial hypersurface $\Sigma(t)$, satisfying
$n_{AA'} n^{AA'}= 1$ and $n_{AA'} e^{AA'}{}_p = 0$. Clearly, $n^{AA'}$
depends on the dynamical variables $e^{AA'}{}_p$; its Dirac brackets are
found to be
\be
[n^{AA'},e^{BB'}{}_q]^\ast = [n^{AA'},\psi^A{}_p]^\ast =
[n^{AA'},\bar\psi^{B'}{}_q]^\ast=0,
\ee
\be
[n^{AA'},p^\pm{}_{BB'}{}^q]^\ast = h^{qr} e^{AA'}{}_r n_{BB'}.
\ee

In addition to the second-class constraints mentioned above, the Lorentz
invariance of the theory gives rise to the first-class constraints
$J_{AB}= \bar J_{A'B'} = 0$ where
\be
\bar{J}_{A'B'}  \equiv  e^A{}_{(A'}{}_p\, p^+{}_{AB')}{}^p \ \ \ \
J_{AB}   \equiv e_{(A}{}^{A'}{}_p\, p^-{}_{B)A'}{}^p. \label{eq:L_charges}
\ee
These quantities can be interpreted as the generators of Lorentz
transformations.

One can identify the charges associated with the other symmetries
of the theory. In particular, the supersymmetry generators are
\be
 \bar{S}_{A'}  \equiv {i\over 2}\kappa^2  \psi^A{}_p p^+{}_{AA'}{}^p\,
\ \ \ \ \ \ \ \ \
S_A \equiv- {i \over 2}\kappa^2  p^-{}_{AA'}{}^p\bar\psi^{A'}{}_p.
\label{eq:SG_charges}
\ee
Indeed, if $f$ is any function of the dynamical variables then its
variation under supersymmetry transformations (\ref{eq:transf}) is
\be
\delta f = - {2\over \kappa} [\epsilon^A S_A + \bar S_{A'}\bar\epsilon^{A'},
f]^\ast.
\label{eq:can_sup}
\ee

Time translations are generated by the Hamiltonian, which is found to be
\be
H= - e^{AA'}{}_0 \H_{AA'} + \psi^A{}{}_0 S_A +  \bar S_{A'} \bar\psi^{A'}{}_0
-\omega^{AB}{}_0 J_{AB} - \bar\omega^{A'B'}{}_0 \bar J_{A'B'}
\ee
where $\omega^{AB}{}_0$ and $\bar\omega^{A'B'}{}_0$ are the $x^0$ components
of the spin connection 1-form \cite{D1}. The expression for $\H_{AA'}$ is quite
complicated, but it is related to the other generators by the identity
\be
\H_{AA'} = - {2i \over \kappa^2} [S_A,\bar S_{A'}]^* \ + \
\hbox{terms proportional to $J$ and $\bar J$}. \label{eq:Ham}
\ee
It can be verified that the classical equations of motion imply the
evolution equation
\begin{equation}
{df \over dt} = [f,H]^*
\end{equation}
for any function $f$ of the dynamical variables.

At this point note that the quantities $e^{AA'}{}_0$, $\psi^A{}_0$ and
$\bar\psi^{A'}{}_0$ are non-dynamical, as their time derivatives do not appear
in the action and consequently the classical equations of motion tell
us nothing about their evolution. If we wish, we may freely specify their
values at all times, thereby eliminating some of the gauge degrees of
freedom.
\bigbreak

\leftline{\bf The Time-Dependent Quantum Theory}

Many authors choose to treat the quantities
$e^{AA'}{}_0$, $\psi^A{}_0$ and $\bar\psi^{A'}{}_0 $ as free variables, i.e.
Lagrange multipliers, and then derive a set of first-class constraints by
requiring that the action be stationary with respect to their variation.
However, one may instead suppose that some of these quantities are fixed
externally by the imposition of gauge conditions; in this case it does not
make sense to require stationarity of the action and so the correponding
constraints are absent from the classical theory.

To quantise the theory, we replace functions of the dynamical variables
by operators and Dirac brackets by commutators or anticommutators.
When the theory is quantised, a function $f$ is then represented
in the Heisenberg picture by an operator which evolves according to
the equation
\begin{equation}
i\hbar {df\over dt} =  [f,H]
\end{equation}
where $[f,H]$ denotes the commutator of two operators. However, we
prefer to use the Schr\"odinger picture, in which the operators
are constant and the wave function $\Psi$ evolves according to the
Schr\"odinger equation
\bea
i\hbar {\partial \over \partial t} \Psi   & = &    H\Psi  \\
& = &  - e^{AA'}{}_0(t) \, \H_{AA'}\Psi  + \psi^A{}{}_0(t) \,S_A\Psi
- \bar\psi^{A'}{}_0(t)  \,   \bar S_{A'}\Psi.
\eea
(The $J,\bar J$ terms have been omitted from the Hamiltonian, since the
wave function is required to be Lorentz invariant.) If one demands that
the wave function should remain independent of $e^{AA'}{}_0$, $\psi^A{}_0$,
$\bar\psi^{A'}{}_0$ (i.e. independent of the choice of gauge) then
differentiation with respect to these quantities leads to the standard
quantum constraints
$\H_{AA'}\Psi = S_A\Psi= \bar S_{A'}\Psi= 0.$

In the present context, however, we want a time-dependent version of the
quantum
theory. This can be achieved by specifying the function $e^{AA'}{}_0(t)$
before quantisation; in other words, by imposing a gauge condition on the
classical theory. Since $e^{AA'}{}_0$ cannot then be viewed as a Lagrange
multiplier, there is no constraint on $\H_{AA'}$ in the classical theory or
on $\H_{AA'}\Psi$ in the quantum theory.

However, fixing $e^{AA'}{}_0(t)$ breaks supersymmetry unless we simultaneously
set
\be
\psi^A{}_0 = \bar\psi^{A'}{}_0= 0.\label{eq:rigid}
\ee
This has the effect of converting the local supersymmetry to a rigid
supersymmetry, since $\epsilon$ and $\bar\epsilon$ must now be constant
so that the conditions (\ref{eq:rigid}) are preserved by supersymmetry
transformations. Note that rigid supersymmetry is required to prove the
existence of Nicolai maps.

With $e^{AA'}{}_0$, $\psi^{A'}{}_0$ and $\bar\psi^{A'}{}_0$ fixed
classically in the manner described above, the wave function will satisfy
the evolution equation
\bea
i\hbar {\partial \Psi \over \partial t}
& = & - e^{AA'}{}_0(t)\, \H_{AA'}\Psi  \\
& = & {2\over \hbar\kappa^2} e^{AA'}{}_0(t)
(S_A \bar S_{A'} + \bar S_{A'} S_A) \Psi. \
\eea
The last line is obtained by using the anticommutation relations
which follow from (\ref{eq:Ham}). We thus have a straightforward
quantum theory with rigid $N=4$ supersummetry.

In order for Nicolai's theorem to be applicable, the quantum state
must be invariant under a non-trivial supersymmetric subalgebra
which preserves the initial and final hypersurfaces.
In fact, we have a choice of possible subalgebras. For definiteness,
suppose that we decide to use the $I_+$ form of the action;
then a suitable subalgebra is the one spanned by
the left-handed supersymmetry generators $\bar S_{A'}$.
The requirement that these correspond to unbroken symmetries leads
to the additional quantum constraints
\be
\bar S_{A'} \Psi = 0 \ \ \ \ \ \ (A'= 0,1)  \label{eq:susy_constraint}
\ee
and so the evolution equation becomes
\be
 {\partial\Psi \over \partial t}  =
- {2i \over \hbar^2 \kappa^2} e^{AA'}{}_0\, \bar S_{A'} S_A\Psi \label{eq:ev}
\ee

There are different operator representations of the quantum theory,
corresponding to the different versions of the classical action.
Having chosen the $I_+$ form of the action, we find that
$e^{AA}{}_p$ and  $\psi^A{}_p$ are the canonical coordinates
and so are represented in the quantum theory by multiplicative
operators. Quantisation is then completed by setting
\be
p^+{}_{AA'}{}^p = -i\hbar{\partial \over \partial e^{AA'}{}_p},\ \ \ \ \ \
{\bar\psi}^{A'}{}_p = - {i\hbar\over\sigma} D^{AA'}{}_{qp}
{\partial\over\partial \psi^A{}_q}
\ee
in accordance with the commutation and anticommutation relations
which follow from the Dirac brackets (\ref{eq:Dir_1}-\ref{eq:Dir_last}).
Operator representations of $\bar J_{A'B'}$ and $\bar S_{A'}$ are
obtained from (\ref{eq:L_charges},\ref{eq:SG_charges}) without any
ordering ambiguities, thanks to (\ref{eq:Dir_2},\ref{eq:Dir_3}).
Moreover, (\ref{eq:new_mom}) gives
\be
p^-{}_{AA'}{}^p = -i\hbar{\partial \over \partial e^{AA'}{}_p}
+   4i \kappa^{-2} \sigma m^{pq}e_{AA'q}
+  \sigma\epsilon^{pqr}[ (1- \factor) \bar\psi_{A'q}\psi_{Ar} -
\factor\psi_{Ar} \bar\psi_{A'q} ]  \label{eq:op_ord}
\ee
where the value of the parameter $\factor$ determines the ordering of
the operators in the $\psi\bar\psi$ term. Note that this is the \it
only \rm operator-ordering ambiguity which will arise.

The supersymmetry generators are now represented by the operators
\begin{equation}
\bar S_{A'}=
 {\hbar\kappa^2\over 2}   {\partial \over \partial e^{AA'}{}_p} \psi^A{}_p,
\label{eq_left.gens}
\end{equation}
\begin{equation}
S_A = - {\hbar\kappa^2\over 2} \bar\psi^{A'}{}_p \left(
{\partial \over \partial e^{AA'}{}_p}
-   {4\sigma\over\hbar \kappa^2} m^{pq}e_{AA'q}
+   2\factor e_{AA'q} h^{qp}\right)
\label{eq_right.gens}
\end{equation}
where $h^{pq}$ is the inverse of $h_{pq}$. Thus, expression (\ref{eq:ev}) gives
\be
{\partial \Psi\over \partial t}  =
i {\kappa^2\over 2} e^{AA'}{}_0\,
{\partial \over \partial e^{BA'}{}_p} \left[  \psi^B{}_p
\bar\psi^{B'}{}_q \left(
{\partial \over \partial e^{AB'}{}_q}
-   {4\sigma\over\hbar \kappa^2} m^{qr}e_{AB'r}
+   2\factor h^{qr} e_{AB'r}\right)\Psi \right]
\label{eq:ferm_evol}
\ee
where the wave function $\Psi$ depends on the canonical
coordinates $e^{AA'}{}_p$ and $\psi^A{}_p$.
\bigbreak

\leftline{\bf The Probability Density Function of the Universe}

A general solution of (\ref{eq:ferm_evol}) is the sum of a number of
distinct components, each of different order in the Grassmann variables
$\psi^A{}_p$. It turns out that certain components are singled
out by the choice of particular boundary conditions.

The Nicolai map will be realised as a stochastic differential equation
relating $\dot e^{AA'}{}_p$ to a white noise process. Consequently, some
kind of boundary conditions on $e^{AA'}{}_p$ are needed to ensure that
this map is invertible. In particular, if we are interested in evolution from a
particular 3-geometry, then we must specify $e^{AA'}{}_p$ at some initial
time. With these boundary conditions, it will turn out that only the
$O(\psi^6)$ component of the wave function is non-zero.

The argument for a probabilistic interpretation depends on the existence
of a Nicolai map, which in this case is assured only if the quantum state
-- and hence any initial conditions --  are invariant under the subalgebra
of left-handed supersymmetry generators $\bar S_{A'}$. In particular, to
ensure that the specified initial values of $e^{AA'}{}_p$ are invariant
under this subalgebra, we must simultaneously impose Dirichlet initial
conditions on the right-handed spinors $\psi^A{}_p$
(since $[ \bar S_{B'},
e^{AA'}{}_p ]^*= {i\over 2 }\kappa^2 \psi^A{}_p \epsilon_{B'}{}^{A'}$).
Together these initial conditions on $e^{AA'}{}_p$ and $\psi^A{}_p$
are $\bar S_{A'}$-invariant, and will give rise to a quantum state
satisfying the constraint (\ref{eq:susy_constraint}).

However, if all the $\psi^A{}_p$ vanish initially, then their equations of
motion imply that they will also vanish at later times. Consequently,
the quantum states arising from the initial conditions described
above will satisfy the six conditions
\be
\psi^A{}_p  \Psi(e, \psi;t)  =0 \ \ \ \ \ \ (p=1,2,3; \ \ \ A=0,1).
\label{eq:ferm_constraint}
\ee
It follows that any $\bar S_{A'}$-invariant states which evolve from
specified initial values of $e^{AA'}{}_p$ must have the form
\be
\Psi(e^{AA'}{}_p,\psi^A{}_p;t) =
(\psi^A{}_p \delta^{pq} \psi_{Aq})^3 \, \F(e^{AA'}{}_p;t)
\label{eq:wf_comp}
\ee
where the function $\F(e^{AA'}{}_p;t)$ is independent of the fermionic
variables. Our argument indicates that these states should admit Nicolai
maps, and therefore probabilistic interpretations.

Making use of the anticommutation relation
$\psi^B{}_p  \bar\psi^{B'}{}_q  =
-{i\hbar\over\sigma} D^{BB'}{}_{pq} - \bar\psi^{B'}{}_q \psi^B{}_p$
(which follows from (\ref{eq:Dir_4})) and using (\ref{eq:ferm_constraint})
we obtain the bosonic evolution equation
\be
{\partial \F \over \partial t}  =  {\kappa^2\over 2\sigma}
e^{AA'}{}_0\,{\partial \over \partial e^{BA'}{}_p}
\left[  D^{BB'}{}_{pq}
\left( \hbar  {\partial \F \over \partial e^{AB'}{}_q }
+ {\partial \Phi \over \partial e^{AB'}{}_q } \F \right) \right]
\label{eq:evol}
\ee
where we have defined
\be
\Phi=   {2\sigma\over\kappa^2} m^{pq}h_{pq} - \factor\hbar \log \det [h_{pq}].
\ee
It is clear from (\ref{eq:evol}) that the integral of $\F(e^{AA'}{}_p;t)$ will
be conserved in time. It can also be shown that (\ref{eq:evol}) is a real
equation when restricted to Euclidean tetrads. (We are interested in
Euclidean spacetimes because only Euclidean theories admit Nicolai
maps with stochastic interpretations).

Given these facts, it is tempting to interpret (\ref{eq:evol}) as a
Fokker-Planck equation describing the evolution of the probability density
function $\F(e^{AA'}{}_p;t)$ for the location of a Brownian particle moving
randomly in a
12-dimensional minisuperspace with coordinates $e^{AA'}{}_p$ and a potential
function $\Phi(e^{AA'}{}_p)$. We will tentatively adopt this intepretation for
now, postponing a more careful investigation until the next section.

Of particular interest is the static solution
\be
\Fh_0(e^{AA'}{}_p) = A \exp -{1\over\hbar} \Phi(e^{AA'}{}_p)
\ee
as this state satisfies all the conventional constraints of homogeneous
supergravity. (It is easily checked that the wave function
$\Psi_0= (\psi^A{}_p \delta^{pq} \psi_{Aq})^3 \Fh_0$ is annihilated by
all four supersymmetry generators (\ref{eq_left.gens},\ref{eq_right.gens})
as well as the Lorentz generators.)

Furthermore, $\Fh_0(e^{AA'}{}_p)$ is a plausible candidate for the
the probability density function of a Universe in the wormhole state.
Indeed, precisely the same p.d.f. can be obtained by squaring the
modulus of the wormhole wave function of Asano \it et al \rm
\cite{ATY}, in accordance with the interpretation advocated by Bene and
Graham \cite{BG}.

A natural question is whether the time-dependent solutions of (\ref{eq:evol})
have any physical significance. According to the traditional view, the answer
is no: One is interested only in the static solution $\Fh_0(e^{AA'}{}_p)$
described above,
since this is in fact the only normalisable state satisifying the standard
constraints of supergravity. From this perspective, it might appear that our
approach merely supports the argument of Bene and Graham \cite{BG} that the
cosmological probability density function should be the square the modulus of
the
conventional wave function.

However it should be noted that our approach, unlike that of Bene and Graham,
provides a plausible probabilistic interpretation of the quantum theory \it
without any reference at all to the concept of an inner product\rm. In this
sense, our approach is radically different from the usual Hilbert space
formulation of quantum mechanics. It thus enjoys a clear advantage in
the arena of quantum cosmology, where the choice of a satisfactory
inner product has long been viewed as one of the most vexing and fundamental
problems.
\bigbreak

\leftline{\bf Time Dependent Solutions}

A more radical approach would be to suppose that the wave
function of the Universe really does evolve with respect to
Euclidean time. The Euclidean nature of time would not necessarily
be apparent to the inhabitants of the Universe, any more than
the time-independence of the cosmological wave function is apparent
to the inhabitants in the standard Wheeler-DeWitt approach.
It might therefore be argued that such an interpretation is quite
compatible with our observations and experience.

To better understand the dynamics implied by the evolution equation
(\ref{eq:evol}), it is convenient to eliminate the nine remaining
gauge degrees of freedom from the 12-dimensional minisuperspace
parametrised by the variables $e^{AA'}{}_p$. (There are six degrees
of freedom associated with Lorentz transformations, and three associated
with the spatial diffeomorphisms corresponding to redefinitions of the
1-forms $\inv^p$.)

To eliminate the Lorentz degrees of freedom, the first step is to
expand the $x^0$ component of the tetrad as
\be
e^{AA'}{}_0 = -i N(t) n^{AA'} + N^p(t) e^{AA'}{}_p
\ee
where $N(t)$ is the Euclidean lapse function and the $N^p(t)$ are the
components of the shift vector. The spacetime metric is then
\be
g = (N^2 +N^p N_p)\dt\otimes\dt  +
N_p (\dt \otimes \inv^p+ \inv^p\otimes\dt)
+ h_{pq} \inv^p \otimes \inv^q
\ee
where $N_p= h_{pq} N^q$.

Since the wave function is invariant under spatial rotations,
its non-vanishing component $\F$ can be written as a function of the
3-metric $h_{pq}$. The evolution equation (\ref{eq:evol})
can then be shown to imply that
\bea
{\partial \Fh \over \partial t}
 & = &   2{\kappa^2\over \sigma}\,  N(t) G^{-{1\over 2}}
{\partial \over \partial h_{pr} }  \left[ G^{1\over 2}\, G_{(pr)(qs)}
\left( \hbar {\partial \Fh \over \partial h_{qs} } +
 {\partial\Phi\over\partial h_{qs} }\Fh   \right) \right] \cr
&  & - 4 N^p(t) {\partial\over \partial h_{qt}}
\left( \epsilon_{pqr} h_{st}m^{rs} \Fh \right)
\label{eq:FP}
\eea
where
\be
G_{(pr)(qs)} ={1\over 2} (\det [h_{pq}] )^{-1/2}\,
( h_{pq}h_{rs} + h_{ps}h_{qr} - h_{pr}h_{qs} )
\ee
is the (contravariant) Wheeler-DeWitt metric, and
$G= {1\over 2} (\det [h_{pq}])^{-1}$ is the absolute value of the
determinant of its inverse \cite{DeWitt}.

Note that the new evolution equation (\ref{eq:FP}) is manifestly real, and
ensures that the integral of $G^{1\over 2} \Fh$ is conserved in time.
This supports the tentative interpretation of $\Fh$ as a probability
density function on the 6-dimensional minisuperspace with coordinates $h_{pq}$.

Using (\ref{eq:FP}) to evolve the wave function forward in time,
we will clearly obtain an expression for $\Fh$ which depends on
the gauge; i.e. on the choice of the $N^p$. On the other
hand, if we do not wish to specify a gauge we should leave these
quantities free as Lagrange multipliers. In this case, the invariance
of the wave function with respect to changes in the $N^p$ leads
to the additonal constraints
\be
0= \epsilon_{trq} h_{sp}m^{qs}{\partial \Fh \over \partial h_{pr}}.
\label{eq:diff_consts}
\ee
When these constraints are imposed, the evolution equation takes the form
\be
{\partial \Fh\over \partial t}   =   {2\kappa^2\over \sigma}\,
N(t)\,  G^{-{1\over 2}}
{\partial \over \partial h_{pr} } \left[ G^{1\over 2}\, G_{(pr)(qs)}
\left( \hbar {\partial\Fh\over \partial h_{qs} } +
 {\partial\Phi\over\partial h_{qs} }\Fh  \right) \right].
\ee

In the Bianchi IX model, the matrix $m^{pq}$ is invertible and so
(\ref{eq:diff_consts}) gives three independent constraints. These
ensure that $\Fh$ is independent of the three remaining gauge degrees
of freedom contained in $h_{pq}$ associated with the possibility of redefining
the invariant 1-forms $\inv^p$ consistently with (\ref{eq:hom_cond}).
Consequently, there are just three
physical degrees of freedom, corresponding to the eigenvalues of the
matrix $Z_p{}^q\equiv h_{pq}m^{qr}$. These eigenvalues are real and positive,
since $Z_p{}^q$ is symmetric and positive-definite in the Bianchi IX case.
In fact these eigenvalues are just the scale factors associated with the three
principle axes of the metric $h_{pq}$.

Suppose that the eigenvalues of $Z_p{}^q$ are arranged in decreasing order and
denoted $e^{2\beta_1},e^{2\beta_2},e^{2\beta_3}$ (so that $\beta_1\geq\beta_2
\geq\beta_3$). Then the physical degrees of freedom of the 3-geometry are most
conveniently parametrised by the variables
\begin{equation}
\alpha = {1\over 3} (\beta_1+ \beta_2 + \beta_3) , \ \ \ \
\beta_+ ={1\over 6} (\beta_1+\beta_2) -  {1\over 3}\beta_3, \ \ \ \
\beta_- ={1\over 2\sqrt{3} } (\beta_1-\beta_2).
\end{equation}
Together, $(\alpha,\beta_+,\beta_-)$ form a coordinate system on the
minisuperspace
of homogeneous $3$-geometries; $\alpha$ is associated
with the overall scale factor, while $\beta_+$ and $\beta_-$ measure the
anistropy
of a given 3-geometry.  Note that our decision to label the eigenvalues in
descending
order means that the coordinates $(\alpha,\beta_+,\beta_-)$ satisfy the
inequalities
\begin{equation}
\beta_+\geq {1\over\sqrt 3} \beta_-  \geq 0
\end{equation}
Note also that the gauge invariance implied by the constraints
(\ref{eq:diff_consts})
implies that the probability density $\Fh$ can be expressed as a function of
the
three minisuperspace coordinates $(\alpha,\beta_+,\beta_-)$ and the time
parameter $t$.

By integrating out the gauge degrees of freedom in the space of 3-metrics, one
finds
that the volume element on the minisuperspace of 3-geometries parametrised by
$(\alpha,\beta_+,\beta_-)$ is $dV =\Omega\,
\d\alpha\wedge\d\beta_+\wedge\d\beta_-$ where
\bea
\Omega(\alpha,\beta_+,\beta_-) &=& e^{(\beta_1+\beta_2+\beta_3)}
(e^{2\beta_1}-e^{2\beta_2}) (e^{2\beta_2}-e^{2\beta_3})
(e^{2\beta_1}-e^{2\beta_3})\cr
&=&  8 e^{9\alpha}\,  \sinh(2{\sqrt 3}\beta_-)\,
\sinh(3\beta_+ -{\sqrt 3}\beta_-)\,\sinh(3\beta_+ + {\sqrt 3}\beta_-)
\eea

The constraints (\ref{eq:diff_consts}) ensure that $\Fh$ is invariant under
gauge transformations and therefore depends only on the physical degrees of
freedom $(\alpha,\beta_+,\beta_-)$ and on $t$. However, instead of dealing
with the combination $\Fh dV $ it is convenient to define
a new probability density function
\begin{equation}
\Fb(\alpha,\beta_+,\beta_-;t)\equiv \Omega(\alpha,\beta_+,\beta_-)\,
\Fh(\alpha,\beta_+,\beta_-;t)
\end{equation}
so that $\Fh dV= \Fb \d\alpha\wedge\d\beta_+\wedge\d\beta_-$. The finiteness of
$\Fh$ then implies that $\Fb$ satisfies the Dirichlet boundary conditions
\begin{equation}
\Fb(\alpha,\beta_+,\beta_-;t) = 0  \ \ \ \
\hbox{if}\ \ \beta_+ =  {1\over\sqrt 3} \beta_- \ \
\hbox{or}\ \ \beta_- = 0.
\end{equation}

{}From (\ref{eq:FP}), it follows that $\Fb(\alpha,\beta_+,\beta_-;t)$ obeys
the evolution equation
\bea
{\partial \Fb\over \partial t} &=&   {\kappa^2 \over 12 \sigma}\, N(t)\,
\left[        {\partial \over\partial \beta_+}
\left( \hbar {\partial \Fb \over\partial \beta_+} +
 {\partial \Phib \over\partial \beta_+} \Fb \right)
\  + {\partial \over\partial \beta_-}
\left( \hbar {\partial \Fb \over\partial \beta_-} +
{\partial \Phib \over\partial \beta_-}\Fb  \right)
\right. \cr
& & \ \ \ \ \ \ \ \ \ \ \ \ \ \ \ \ \ \ \ \ \ \ \ \ \ \ \left.
 -{\partial \over\partial \alpha}
\left( \hbar {\partial \Fb \over\partial \alpha} +
{\partial \Phib \over\partial \alpha}\Fb  \right)
\right]  \label{eq:evol2}
\eea
where we have introduced the modified potential $\Phib=\Phi-\hbar\log \Omega$,
which
in terms of the coordinates $\alpha,\beta_+,\beta_-$ has the form
\bea
\Phib &=& {2\sigma\over \kappa^2} e^{2\alpha}
\left[ 2e^{2\beta_+} \cosh (\sqrt{12}\beta_-) + e^{-4\beta_-} \right]
- \hbar (9+\factor)\alpha  - \hbar\log \sinh(\sqrt{12}\beta_-) \cr
& & - \hbar\log \sinh(3\beta_+ -{\sqrt 3}\beta_-)
- \hbar\log \sinh(3\beta_+ + {\sqrt 3}\beta_-).
\eea

The evolution equation (\ref{eq:evol2}) is manifestly real, and
ensures that the integral of $\Fb$ is constant. It has the form of a
conventional three-dimensional Fokker-Planck equation, but with one unusual
feature; the $-$ sign preceeding the $(\partial/\partial\alpha)^2$ term
indicates that the diffusion matrix is not positive-definite.
Equations of this kind are known sometimes as ``pseudo-Fokker-Planck"
equations, and do not always admit normalisable non-singular
solutions when subject to boundary conditions of the kind
discussed above \cite{FPE}.
(However, it should be noted that (\ref{eq:evol}) has at least one well-behaved
solution; namely, the static and supersymmetric solution
$\Fb_0= \Omega\Fh_0= \exp (-U/\hbar)$.)

{}From a geometric point of view, the problem arises because the minisuperspace
metric has a Lorentzian signature ($-++$). This is a direct consequence
of the fact that the Euclidean Einstein-Hilbert action is not
positive definite \cite{H}; analogous difficulties arise in all
approaches to the quantisation of gravity. In the path-integral prescription,
these problems are often dealt with by analytic continuation, with the
contour of integration for the conformal part of the metric being rotated
in the complex plane \cite{H}.

In the present case, the non-positive-definiteness of the diffusion matrix
means that normalisable solutions to (\ref{eq:evol}) will generally develop
singularities after a finite time. This precludes a naive intepretation
of $\Fb(\alpha,\beta_+,\beta_-;t)$ as a time-dependent probability density
function. Recall, however, that the primary use of probabilities is for
calculating expectation values; a very natural approach to the problem is
therefore to analytically continue the definition of the expectation value.
(Note that analytic continuation has been applied successfully
in quite different contexts to make sense of probabilities which are
formally defined as integrals of non-normalisable density functions.
For example, see Appendix C of ref. \cite{LM}.)

To see how this approach might work, suppose that
$P(\alpha,\beta_+,\beta_-;t)$ is some un-normalised probability density
function for the random variables $\alpha(t),\beta_+(t),\beta_-(t)$ at a
given time $t$. Then the expectation value at this time of a quantity $Q$
which depends on $\alpha,\beta_+, \beta_-$ is given by the ratio
\begin{equation}
E[Q] = { \int d\alpha \int d\beta_+ \int d\beta_- \,
Q(\alpha,\beta_+,\beta_-)\, \Fb (\alpha,\beta_+,\beta_-;t)\,
\over  \int d\alpha \int d\beta_+ \int d\beta_- \, \Fb
(\alpha,\beta_+,\beta_-;t) }
\label{eq:expec}
\end{equation}
where the contours of integration are along the real axes.

However, if $P(\alpha,\beta_+,\beta_-;t)$ is a solution to (\ref{eq:evol})
evolving from typical initial data then this simple definition of the
expectation
value may break down. Indeed, if $\Fb$ is initially supported inside a small
region of $(\alpha,\beta_+,\beta_-)$-space, then it will quickly develop
singularities at points along the real $\alpha$ axis. It will then become
impossible to evaluate the integrals appearing in (\ref{eq:expec}), and so
this expression will fail to give a well-defined expectation value.
A natural way to avoid this problem is by shifting the contour of the
$\alpha$-integration in (\ref{eq:expec}) away from the real axis so that the
singularities are avoided. Hopefully a contour $\C$ can be found which gives
rise to expectation values with all the standard properties.

Certainly, any such analytically continued expectation value $E[Q]$ will
depend linearly on $Q$, in accordance with the standard rules of probability
theory. Moreover, provided that the contour $\C$ is symmetric
under reflection in the real axis, $E[Q]$ will be real for any function
$Q(\alpha,\beta_+,\beta_-)$ which takes real values along the real
$\alpha$-axis. The only remaining requirement for bona-fide expectation
values is that, if the function $Q(\alpha,\beta_+,\beta_-)$ is positive for
all real $\alpha,\beta_+,\beta_-$, then $E[Q]$ should also be positive.
At present, it is not clear whether a contour can be found to
satisfy this third criterion.

If an appropriate contour $\C$ exists, then this prescription
leads to a satisfactory probabilistic intepretation even for solutions
of (\ref{eq:evol}) which are non-normalisable or develop singularities
on the real $\alpha$ axis. A similar type of analytic continuation was
used in \cite{GL1}, where the contour $\C$ was rotated to the imaginary
$\alpha$ axis.

On the other hand, it is possible that a satisfactory contour may
not exist. This would mean that, whatever analytic continuation scheme was
employed, negative results might sometimes be obtained when evaluating
expectation values of positive quantities. This would contradict the usual
rules of probability theory, and thus call into question the feasibility of
a probabilistic intepretation for the time-dependent states.

At present, it is not clear whether a satisfactory analytic continuation
procedure can be found, or if there is any other way in which (\ref{eq:evol2})
can be converted into the evolution equation for a bona-fide probability
density function. Consequently, the physical significance of the
time-dependent solutions remains uncertain.

\bigbreak

\leftline{\bf Discussion}

We have seen how, by imposing certain gauge and coordinate conditions on
the classical theory, the quantisation of homogeneous (Euclidean)
supergravity leads to an evolution equation for the wave function
which closely resembles a conventional Fokker-Planck equation.

However, because the Einstein-Hilbert action is unbounded below,
the diffusion matrix is not positive definite and consequently the
class of non-singular normalisable solutions is drastically reduced.
In fact it is quite possible that there are no well-behaved
time-dependent solutions at all, which raises interesting questions
about the nature of the supposed evolution.

Nonetheless, the evolution equation (\ref{eq:evol2}) is known to admit
a well-behaved static solution, which can be consistently intepreted as a
cosmological probability density function. This solution has the form
$\Fb_0= \exp (- U/\hbar)$, and is known to satisfy all the standard
quantum constraints for the theory. It is encouraging that
this probability density function coincides with that obtained for
the wormhole state by more conventional approaches to quantum the
theory.

An important question is whether the Hartle-Hawking state
can, like the wormhole state, be interpreted in the manner described
above. At present the answer to this question is unclear. It is known
that an alternative definition of homogeneity must be used if one
wants to find such a state in a minisuperspace theory \cite{GL2};
however, the imposition of this alternative ansatz on the fields
may prevent them from obeying their classical equations of motion.
Thus, classical trajectories in the minisuperspace model might not
correspond to classical trajectories in the full theory. Clearly,
the absence of such a correspondence would raise grave doubts about
the relevance of minisuperspace as a tool for modelling the dynamics
of the full theory.

In general, one would hope eventually to go beyond the strictures of
minisuperspace. In this context, it should be emphasised that the
approach outlined above should be applicable to a very wide range
of supersymmetric models; the only requirement is that the action
can be made precisely invariant under a boundary-preserving
subalgebra of supersymmetry transformations.

It is unfortunate that, without the imposition of homogeneity
conditions, pure $N=1$ supergravity fails this test.
However this failure cannot be viewed as a serious limitation
on the applicability of our approach, since supergravity
also fails the more elementary test of consistency in the
finite fermion sector \cite{CFOP}. Indeed, it seems likely
that a fully consistent supersymmetric theory \it would \rm
be amenable to the approach described above, without the
need for reduction to minisuperspace. We hope to investigate
this question in future work.

\bigskip
\bigskip
\bigbreak

\leftline{\bf Acknowledgements}
H.L. is grateful to Stephen Poletti and Chris Cosgrove for helpful and
illuminating discussions.
\bigbreak

\vfill
\end{document}